\documentclass[sort&compress,5p]{elsarticle}
\usepackage{graphicx,natbib,amssymb}
\usepackage[pdftex,colorlinks,citecolor=blue,bookmarks]{hyperref}
\usepackage{bm}
\usepackage{amsmath}
\begin{document}

\title{Correlations and neutrinoless $\beta\beta$ decay nuclear matrix elements of $pf$-shell nuclei}
\author[TU,EMMI]{Javier Men\'endez}
\author[TU]{Tom\'as R. Rodr\'iguez}
\author[TU,GSI]{Gabriel Mart\'inez-Pinedo}
\author[UAM]{Alfredo Poves}

\address[TU]{Institut f\"ur Kernphysik, Technische Universit\"at Darmstadt, Schlossgartenstr. 2, D-64289 Darmstadt, Germany}
\address[EMMI]{ExtreMe Matter Institute EMMI, GSI Helmholtzzentrum f\"ur Schwerionenforschung GmbH, D-64291 Darmstadt, Germany}
\address[GSI]{GSI Helmholtzzentrum f\"ur Schwerionenforschung, Plankstr. 1, D-64291 Darmstadt, Germany}
\address[UAM]{Departamento de F\'isica Te\'orica and IFT UAM/CSIC, Universidad Aut\'onoma de Madrid, Cantoblanco, E-28049, Madrid, Spain}

\begin{abstract}
We calculate the nuclear matrix elements (NMEs) for neutrinoless double-beta decays ($0\nu\beta\beta$) of $pf$-shell nuclei
using the shell model (SM) and energy density functional (EDF) methods.
The systematic study of non-physical decays (except for $^{48}$Ca) of Ca$\rightarrow$Ti, Ti$\rightarrow$Cr and Cr$\rightarrow$Fe allows for a detailed comparison between the two nuclear structure approaches.
We observe that while the dominant Gamow-Teller part of the NME differs roughly by a factor of two between SM and EDF,
when we restrict the calculations to spherical EDF states and seniority-zero SM configurations, the NMEs obtained by both methods are strikingly similar.
This points out to the important role of nuclear structure correlations for $0\nu\beta\beta$ decay NMEs.
We identify correlations associated with high-seniority components in the initial and final states of the decay as one of the
reasons for the discrepancies between SM and EDF results.
We also explore exact projection to good isospin, and conclude that it has only a moderate effect in the Gamow-Teller part of the NMEs but strongly affects the Fermi contribution.
This work opens up the door for NME benchmarks between different theoretical approaches, and constitutes a step forward towards more reliable estimations of the NMEs. 
\end{abstract}
\maketitle
\section{Introduction}~\label{intro}
Experimental searches for the lepton-number violating weak process neutrinoless double beta decay ($0\nu\beta\beta$ decay)
are the most promising approach to determine some of the fundamental properties of neutrinos.
The detection of $0\nu\beta\beta$ decay would establish the Majorana character of neutrinos and provide information about their absolute mass
and hierarchy~\cite{RMP_80_481_2008}.
Ongoing experiments EXO~\cite{exo}, KamLAND-Zen~\cite{kamland} and GERDA~\cite{gerda}
have recently set impressive lower-limits, well over $10^{25}$ years, on the half-lives of $^{136}$Xe and $^{76}$Ge,
and the dozen most favorable nuclei are being explored worldwide 
~\cite{EPJC_73_2330_2013,JPCS_381_012044_2012,JPCS_375_042018_2012,PPNP_64_267_2010,AIPCP_942_101_2007,PRC_80_032501_2009,JPCS_365_042023_2012,PRC_78_035502_2008,JINST_8_P04002_2013}. 
However, an eventual $0\nu\beta\beta$ decay detection does not guarantee the precise determination of absolute neutrino masses,
because the half-life depends on the transition nuclear matrix elements (NME),  $M^{0\nu}$~\cite{RMP_80_481_2008}:
\begin{equation}
\left[T_{1/2}^{0\nu}(0_{i}^{+}\rightarrow0_{f}^{+})\right]^{-1}=G_{0\nu}|M^{0\nu}|^{2}\left(\frac{\langle
    m_{\beta\beta}\rangle}{m_{e}}\right)^{2}, 
\label{half_life}
\end{equation} 
with $m_e$ the electron mass and $G_{0\nu}$ a well-known phase-space factor~\cite{PRC_85_034316_2012, Stoica_phasespace}.
The combination of neutrino masses that appears in $0\nu\beta\beta$ decay is $\langle m_{\beta\beta}\rangle=|\sum_{i} U_{ei}^2 m_i|$,
with $U_{ij}$ the neutrino mixing matrix.

NMEs have been predicted by different theoretical nuclear structure approaches.
These comprise large-scale shell model (SM)~\cite{PRL_100_052503_2008,NPA_818_139_2009},
energy density functional methods (EDF)~\cite{PRL_105_252503_2010,Nuria_PRL},
the quasiparticle random phase approximation (QRPA)~\cite{Simkovic_isospin, suhonen_jpg},
and the interacting boson model (IBM)~\cite{Iachello_2013}.
However, state-of-the-art NME predictions by these approaches differ up to factor two~\cite{vogel},
strongly limiting the precision to which information on neutrino masses would be known in case of a $0\nu\beta\beta$ decay measurement.
In addition, the half-lives of experimentally relevant nuclei could be significantly under/overestimated for a given $\langle m_{\beta\beta}\rangle$.

A better understanding of the NMEs is therefore crucial.
Recently, experimental observables relevant for $0\nu\beta\beta$ decay have been measured~\cite{Schiffer_jpg,Schiffer_130,Frekers},
allowing checks for the nuclear structure methods~\cite{Alfredo_2nu},
that in some cases resulted in closer NMEs between different theoretical approaches~\cite{PRC_79_015502_2009,prc_A76,NPA_847_207_2010}.

In this article we follow a complementary approach,
studying non-physical $0\nu\beta\beta$ decays along isotopic chains in the $pf$-shell, with SM and EDF methods.
Although within this region only $^{48}$Ca is an actual $0\nu\beta\beta$ decay candidate,
a comparison between methods is better established with systematic calculations.
For instance, systematic EDF calculations for the cadmium isotopes
provided a better understanding of the role of deformation, pairing and shell effects in $0\nu\beta\beta$ decay~\cite{PLB_719_174_2013}. 
Systematic studies allow to analyze not only numerical values, but also trends,
which are useful to identify similarities and differences between SM and EDF. 
Understanding these is essential to reduce the theoretical uncertainty in the $0\nu\beta\beta$ decay NMEs.

\section{Theoretical framework}~\label{theo_frame}
Here we briefly describe the EDF and SM calculations performed in this work,
as well as the $0\nu\beta\beta$ decay transition operator employed.
A more extensive description of EDF calculations can be found in Ref.~\cite{PRL_105_252503_2010} and references therein.
SM details can be found in Refs~\cite{SMRMP,NPA_818_139_2009}.
The transition operator is discussed in depth in Refs.~\cite{PRC_60_055502_1999,NPA_818_139_2009}.

\subsection{Energy density functional}
The initial and final EDF states are found using the Gogny D1S functional~\cite{NPA_428_23_1984}. Beyond mean field effects such as particle number and angular momentum restoration are included in addition to axial quadrupole configuration mixing within the generator coordinate method (GCM)~\cite{RingSchuck}. 
This method has been extensively used to study several nuclear structure properties throughout the whole nuclear chart
(see for instance Ref.~\cite{PRC_81_014303_2010} for its global performance with an approximate GCM method).
One of the advantages of this approach is the explicit calculation of the NMEs as a function of the quadrupole deformation of initial and final nuclei.
Hence, full shape-mixing and spherical NMEs can be compared. 
Additional degrees of freedom such as fluctuations in the pairing field (already applied to NMEs in Ref.~\cite{Nuria_PRL}),
triaxiality or octupolarity are neglected due to prohibitive computational times.
Nevertheless, we do not expect from these improvements any qualitative difference with respect to the analysis presented in this work. 

\subsection{Shell model}
SM calculations are performed in the valence space comprised by the 0$f_{7/2}$, 1$p_{3/2}$, 1$p_{1/2}$ and 0$f_{5/2}$ orbitals ($pf$ shell),
on top of a $^{40}$Ca core.
We employ the well-known KB3G~\cite{KB3G} and GXPF1A~\cite{GXPF1} effective interactions,
which have been shown to describe well the nuclear structure of $pf$ shell nuclei~\cite{SMRMP}.
While the configuration space is significantly smaller in SM than in EDF calculations,
the main advantage of the SM approach is that all possible correlations within this space are included.
The effect of the reduced SM valence space has been recently studied in the framework of many-body perturbation theory~\cite{EngelHolt,48Qvalue},
with moderate increases in the NMEs. Here we neglect these corrections.

Truncated calculations can be performed limiting the number of neutrons and protons
not coupled in $J=0$ pairs (seniority truncations)~\cite{PRL_100_052503_2008},
enabling the study of the correlations associated to high-seniority components.
In addition, SM states obtained in the full $pf$ shell have good isospin quantum number,
and projection to good isospin can be performed for truncated calculations.
The SM code NATHAN~\cite{SMRMP} has been used throughout this work.

\subsection{Nuclear Matrix Elements}
With the initial and final states obtained with the EDF and SM methods, we calculate the $0\nu\beta\beta$ decay NMEs as described in Refs.~\cite{PRL_105_252503_2010,NPA_818_139_2009}.
These can be decomposed according to spin structure into three different terms
\begin{equation}
M^{0\nu}=M^{0\nu}_{GT}-\left(\frac{g_{V}}{g_{A}}\right)^{2}M^{0\nu}_{F} -M^{0\nu}_{T},
\label{MMM}
\end{equation}
\vspace{-3mm} \\
where $g_{V}=1.0$ and $g_A=1.25$ are the vector and axial coupling constants, respectively.
The Gamow-Teller (GT) part, $M^{0\nu}_{GT}$, is dominant,
and the Fermi (F) component, $M^{0\nu}_{F}$, accounts to $10\%-35\%$ of $M^{0\nu}_{GT}$~\cite{NPA_818_139_2009, Simkovic_isospin,Nuria_PRL,Iachello_2013}.
The tensor contribution, $M^{0\nu}_{T}$, gives a very small correction~\cite{NPA_818_139_2009,PRC_75_051303_2007}.
Here we focus on the main NME component $M^{0\nu}_{GT}$, and discuss $M^{0\nu}_{F}$ in the context of isospin conservation.

We assume the closure approximation, 
which has been shown to be a good approximation (up to 10\%)~\cite{PRC_83_015502_2011, Horoi_closure},
sufficient for the purposes of this work.
In this scheme the GT and F parts of the NMEs follow the transition operator evaluated between the initial and final states:
\begin{equation}
M^{0\nu}_{F/GT}=\langle
0^{+}_{f}|\hat{M}^{0\nu}_{F/GT}|0^{+}_{i}\rangle\label{NME_EV},
\end{equation} 
\vspace{-5mm} \\
with
\begin{eqnarray}
  \label{eq:1}
  \hat{M}^{0\nu}_{F}& = &\left(\frac{g_{A}}{g_{V}}\right)^{2}\sum_{i<j} V_{F}(r_{ij},A,\mu)
  \,\tau_{i}^{-}\tau_{j}^{-}, \\
  \hat{M}^{0\nu}_{GT} & = & \sum_{i<j} V_{GT}(r_{ij},A,\mu)
  \,\bm{\sigma}_{i} \cdot \bm{\sigma}_{j}\,\tau_{i}^{-}\tau_{j}^{-},     
\end{eqnarray}
\vspace{-3mm} \\
where $\tau^{-}$ is the isospin-lowering operator
that transforms neutrons into protons, and $\bm{\sigma}$ are the Pauli spin matrices.
The neutrino potentials $V_{F/GT}$ depend on the relative distance between the two decaying nucleons, $r_{ij}$,
the mass number $A$, and the closure energy $\mu$. Here we use $\mu=7.72$ MeV for all decays,
the standard value used for $^{48}$Ca~\cite{PRL_100_052503_2008,PRL_105_252503_2010}.
A detailed form of the neutrino potentials can be found in Refs.~\cite{PRC_60_055502_1999,NPA_818_139_2009}.

In addition, short-range correlations are included within the UCOM framework~\cite{SuhonenUCOM,NPA_632_61_1998}. 
Other prescriptions 
have been recently proposed~\cite{MutherSimkovic_SRC},
but the differences are small, and not relevant for the purpose of this work.

Here we neglect two-body currents, related to the effective quenching of the $\bm{\sigma}\tau$ operator~\cite{MGS2011} in weak decays,
and restrict to a purely two-body operator derived from one-body currents only.
While two-body current contributions may be important for the absolute value of the NMEs~\cite{MGS2011}, including them would not alter the main conclusions of this study.

\begin{figure*}[t]
  \includegraphics[width=\textwidth]{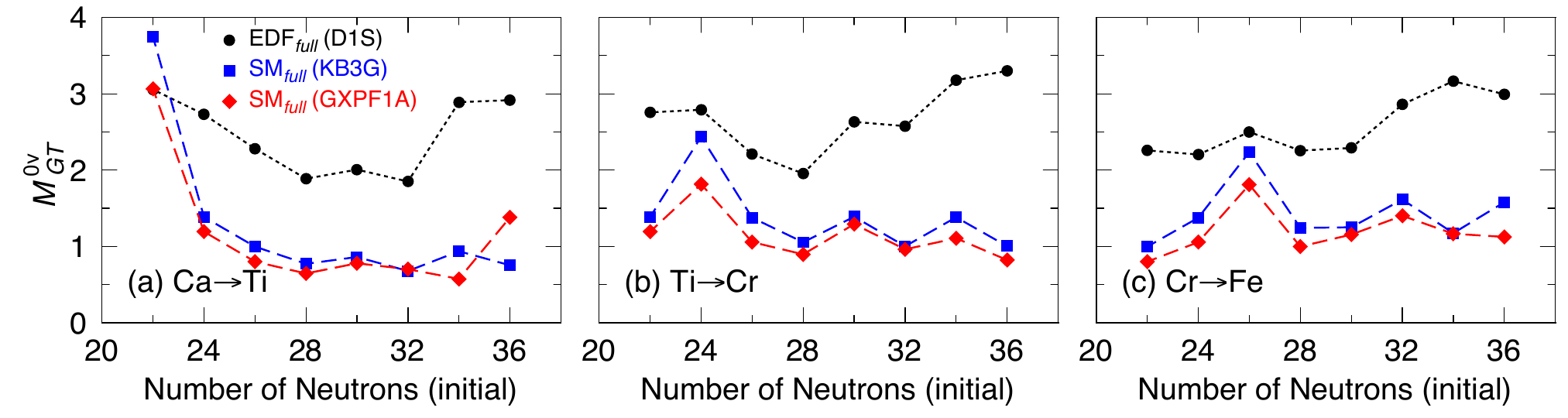}
\caption{(color online)
Gamow-Teller part of the nuclear matrix element, $M^{0\nu}_{GT}$,
for Ca$\rightarrow$Ti (panel a), Ti$\rightarrow$Cr (panel b) and Cr$\rightarrow$Fe (panel c) non-physical $0\nu\beta\beta$ decays,
calculated with shell model (SM) and energy density functional (EDF) methods. 
The D1S EDF interaction is used (circles).
In the SM case, the KB3G (squares) and GXPF1A (lozenges) effective interactions are employed.
}\label{full_nme} 
\end{figure*}
\begin{figure*}[t]
  \includegraphics[width=\textwidth]{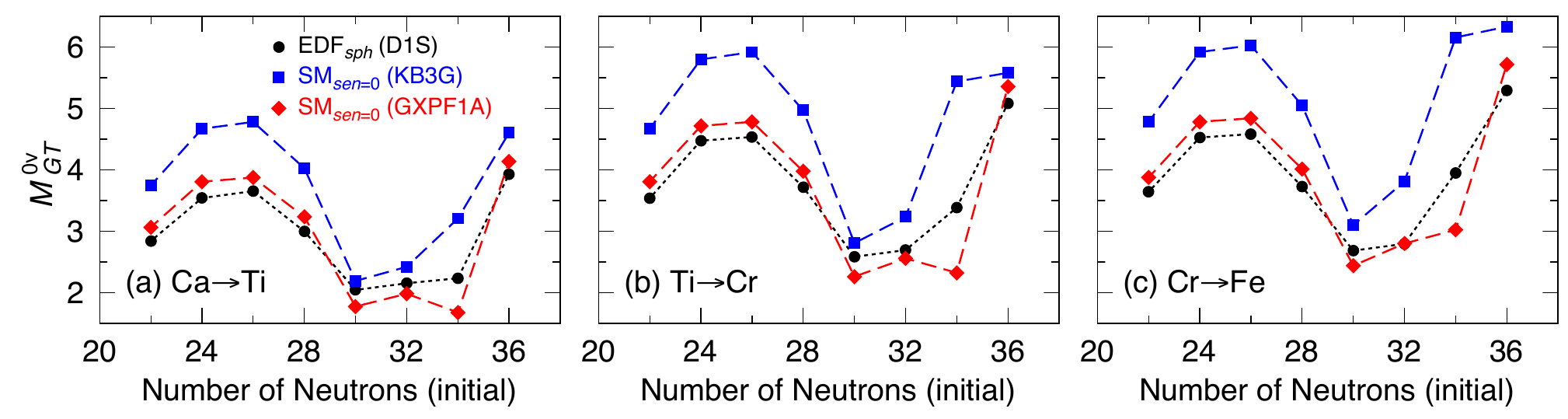}
\caption{(color online)
Gamow-Teller part of the nuclear matrix element, $M^{0\nu}_{GT}$,
for Ca$\rightarrow$Ti (a), Ti$\rightarrow$Cr (b) and Cr$\rightarrow$Fe (c) non-physical $0\nu\beta\beta$ decays,
with seniority-zero shell model (SM) and spherical energy density functional (EDF) states. 
Interactions are as in Fig. \ref{full_nme}.
}\label{sph_nme} 
\end{figure*}

\section{Results}~\label{results}

With the transition operator described in Sec.~\ref{theo_frame}, identical for SM and EDF calculations,
we can make a direct comparison between the NMEs obtained by both approaches.
We have calculated the NMEs for the $0\nu\beta\beta$ decay of the Ca, Ti and Cr isotopic chains.
The GT parts of the NMEs are compared in Fig.~\ref{full_nme} for SM and EDF calculations.
As in the case of actual $0\nu\beta\beta$ decay candidates~\cite{PRL_105_252503_2010,NPA_818_139_2009},
the SM NMEs are about half of the EDF values.
Moreover, this difference is independent on the particular interaction used. 
For the SM, results with two effective interactions, KB3G and GXPF1A, are shown,
with differences of around $10\%-20\%$. This agrees with and extends previous studies in the pf shell restricted to $^{48}$Ca~\cite{EPJA_Poves,HoroiStoica}.
For the EDF, we have also calculated NMEs with the Gogny D1M functional, which results in very small differences with respect to the Gogny D1S.

Figure~\ref{full_nme} also reveals that the trends along the isotopic chain are similar in SM and EDF calculations.
In particular, relative maxima are found in the decays of mirror nuclei: $^{42}$Ca$\rightarrow^{42}$Ti, $^{46}$Ti$\rightarrow^{46}$Cr and $^{50}$Cr$\rightarrow^{50}$Fe,
in agreement with Refs.~\cite{inpc_poves, PLB_719_174_2013}.
Maxima are more marked in SM calculations, where the initial and final states share the same isospin quantum number, $T$.
In the SM case the two states are exactly isospin-symmetric, because Coulomb and isospin-symmetry-breaking terms in the nuclear interaction are neglected,
but the overlap between mirror initial and final states is also maximal in the EDF approach, which includes the Coulomb term.
For EDF calculations, however, $T$ is not a good quantum number.

The configuration space and nuclear correlations included in SM and EDF calculations are very different,
with the SM being able to take into account more general correlations but in a rather limited valence space.
Regarding the size of the configuration space it is important to note that in the pf shell the SM includes all orbitals with their corresponding spin-orbit partner.
This is relevant because in the $0\nu\beta\beta$ decay of heavier nuclei, some spin-orbit partners are not included in SM calculations,
and this has been pointed out as a possible cause of the relatively small SM NMEs.
The SM calculations analyzed in this work are thus free from this shortcoming.
\begin{figure*}[t]
  \includegraphics[width=\textwidth]{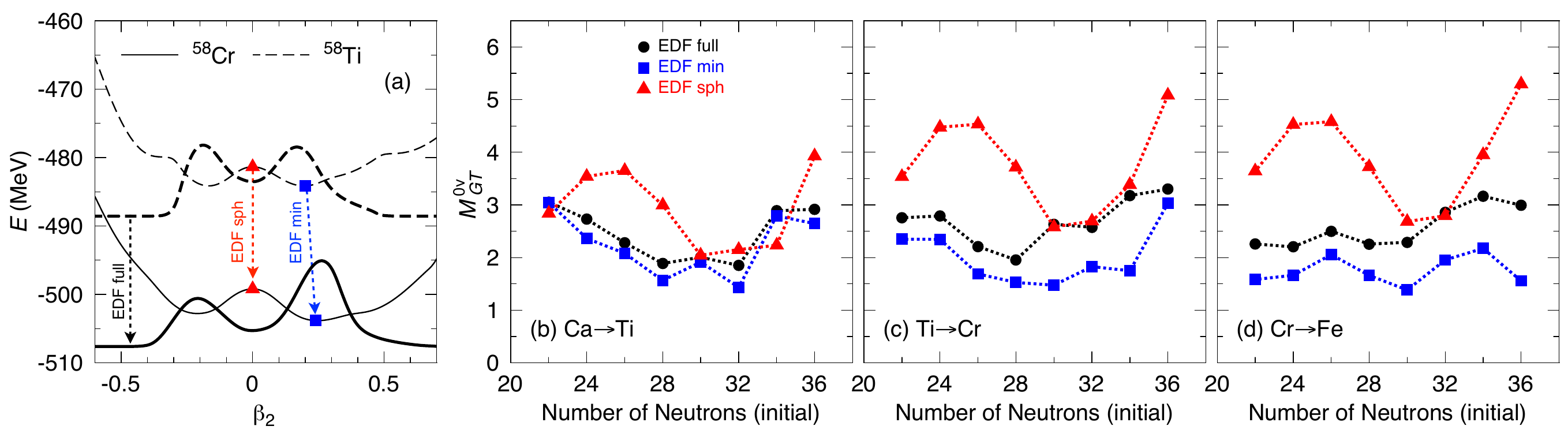}
\caption{(color online)
(a) Particle-number and angular-momentum projected ($J=0$) potential energy surfaces (thin lines) and ground-state collective wave functions (thick lines)
for $^{58}$Ti (dashed) and $^{58}$Cr (solid) nuclei as a function of the quadrupole deformation $\beta_{2}$.
Triangles (squares) correspond to the spherical points (minima) of the corresponding surfaces.
(b)-(d) Gamow-Teller part of the nuclear matrix element, $M^{0\nu}_{GT}$,
for (b) Ca$\rightarrow$Ti, (c) Ti$\rightarrow$Cr and (d) Cr$\rightarrow$Fe (d) non-physical $0\nu\beta\beta$ decays. Calculations use the Gogny D1S functional with initial and final states obtained at the level of spherical calculation (red triangles),
taking the minimum of the potential-energy surface in a deformed calculation (blue squares), and in the full calculation with configuration mixing (black circles) -see panel (a) for the different approaches.}\label{Fig3_EDF} 
\end{figure*}

We can get more insight in the comparison of SM and EDF NMEs by simplifying the nuclear structure correlations present in the initial and final states of the $0\nu\beta\beta$ decay.
Figure~\ref{sph_nme} shows $M^{0\nu}_{GT}$ calculated with the same transition operator as Fig.~\ref{full_nme}, but with simplified nuclear states.
For the EDF, spherical symmetry is assumed.
In the SM case, only configurations with zero seniority ($s=0$) are permitted, this is, protons and neutrons are coupled in $J=0$ pairs -no proton-neutron $J=0$ pairs are included.
We observe that the GT parts of the NMEs calculated in these simplified schemes are significantly larger than in the full calculation for both approaches,
with an striking agreement between SM and EDF NMEs. Indeed SM GXPF1A calculations lie within $10\%$ from EDF values, while SM KB3G calculations are about $25\%$ larger.
The difference between the two SM results stems from the different $J=0$, $T=1$ pairing. 
As shown in Fig.~\ref{full_nme}, this difference between effective interactions is washed out when full calculations are performed.
The agreement between SM and EDF NMEs is in strong contrast with the full NME calculations shown in Fig.~\ref{full_nme}, where SM NMEs were half of the EDF values.

This implies that the spherical EDF and seniority-zero SM calculations, while conceptually very different,
capture approximately the same physics, leaving out the nuclear structure correlations that reduce the $0\nu\beta\beta$ decay NMEs.
These have been identified in Refs.~\cite{PRL_100_052503_2008, PRL_105_252503_2010, PLB_719_174_2013}
as the correlations associated with high-seniority components in the SM, and collective deformation effects in EDF calculations.

Figure~\ref{sph_nme} also shows that the trends followed by the NMEs calculated in both approaches are very similar,
and indeed they follow to a good approximation the generalized seniority scheme in a single shell~\cite{PRC_79_044301_2009}:
\begin{equation}
M^{0\nu}_{GT}\simeq \alpha_{\pi}\alpha_{\nu}\sqrt{N_{\pi}+1}\sqrt{\Omega_{\pi}-N_{\pi}}\sqrt{N_{\nu}}\sqrt{\Omega_{\nu}-N_{\nu}+1}
\label{GS}
\end{equation}
where $N_{\pi(\nu)}$ is the number of proton (neutron) pairs in the shell, $\Omega_{\pi(\nu)}$ the pair degeneracy and $\alpha_{\pi(\nu)}$ coefficients characteristic of a major shell.
Deviations from Eq.~(\ref{GS}) are due to non-perfect shell closures and the $A$ dependence in the neutrino potentials.
The ``inverted parabola'' from initial number of neutrons $N_i=22$ to $N_i=30$, common to all cases, shows the filling of the neutron $f_{7/2}$ orbital associated to
the shell closure at neutron number $N=28$.
The rather ``flat'' behavior between $N_i=30$ to $N_i=32$ is governed by the filling of the $p_{3/2}$ orbital, associated with the closure at $N=32$.
At this point, NMEs obtained with the SM GXPF1A interaction decrease, due to the filling of the $p_{1/2}$ orbital associated with the $N=34$ closure,
while the SM KB3G and EDF results, which do not predict such a closure, increase.
In all cases the NMEs at $N_i=36$ are larger because the $f_{5/2}$ orbital is starting to get filled.
Furthermore, Eq.~(\ref{GS}) predicts that due to the filling of the proton $f_{7/2}$ shell ($\Omega_{\pi}$=4),
the NMEs for Ti and Cr decays ($N_{\pi}=1,2$) to be the same, as observed in Fig.~\ref{sph_nme}.

The fact that both seniority-zero SM and spherical EDF calculations agree with the generalized seniority scheme,
and result in very similar NMEs, opens up the door to benchmarking also NMEs calculated with other nuclear structure methods,
such as QRPA or IBM, which using similarly simplified initial and final states should also agree with the results of Fig.~\ref{sph_nme}.

\begin{figure}[t]
  \includegraphics[width=\columnwidth]{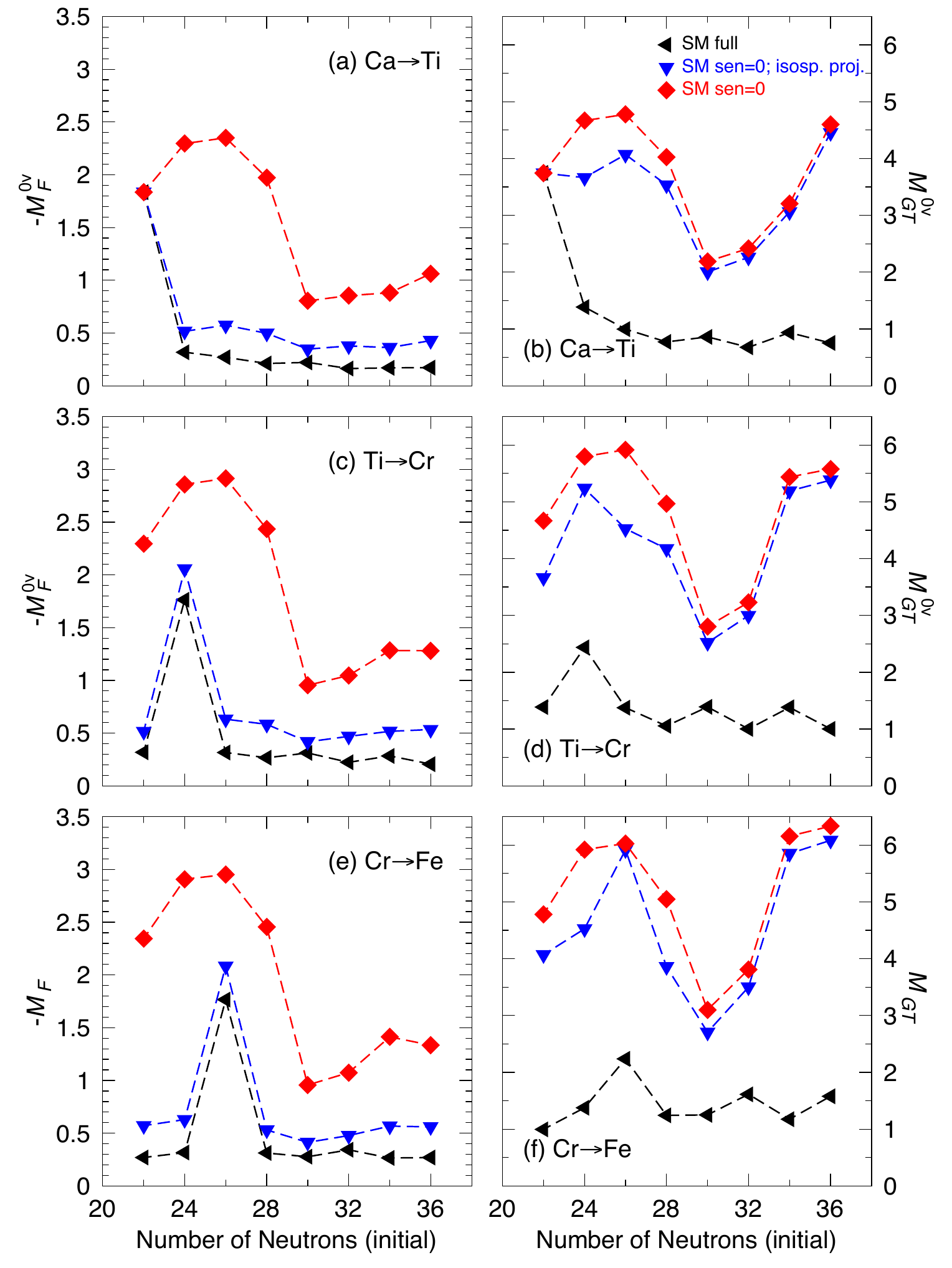}
\caption{(color online)
Shell Model results for the Fermi (left panels) and Gamow-Teller (right panels) parts of the nuclear matrix element, $M^{0\nu}_F$ and $M^{0\nu}_{GT}$, for Ca$\rightarrow$Ti (a)-(b), Ti$\rightarrow$Cr (c)-(d) and Cr$\rightarrow$Fe (e)-(f) $0\nu\beta\beta$ decays. Calculations are performed -using KB3G interaction- with initial and final states obtained at zero-seniority (red diamonds),
zero-seniority with exact isospin projection (blue inverted triangles) and in the full calculation (black left triangles).}\label{Fig3_SM} 
\end{figure}

The role of nuclear structure correlations in $M^{0\nu}_{GT}$ is studied in Figs.~\ref{Fig3_EDF}-~\ref{Fig3_SM},
where the full EDF and SM results of Fig.~\ref{full_nme} are compared to the spherical EDF and seniority-zero SM results of Fig.~\ref{sph_nme}.
In addition, Figs.~\ref{Fig3_EDF}-~\ref{Fig3_SM} also show intermediate results that give additional information on the role of correlations into $0\nu\beta\beta$ decay NMEs.

Within the EDF approach we can explore the intrinsic quadrupole deformation $\beta_{2}$ in the initial and final nuclei, as well as the effect of shape mixing.
Fig.~\ref{Fig3_EDF}(a) shows potential energy surfaces (PES, thin lines) projected to particle-number and angular momentum for $^{58}$Ti$\rightarrow^{58}$Cr,
and the ground-state collective initial and final states obtained after configuration mixing (thick lines).
Fig.~\ref{Fig3_EDF}(a) distinguishes the three EDF calculations in Fig.~\ref{Fig3_EDF}, panels (b), (c) and (d).
In the spherical calculation (EDF$_{\text{sph}}$ in Fig.~\ref{Fig3_EDF})
the initial and final states are the spherical $\beta_{2}=0$ states denoted with triangles in Fig.~\ref{Fig3_EDF}(a).
A better approach consists in considering the minima of the corresponding PES to calculate the NMEs (EDF$_{\text{min}}$). 
Finally, the full EDF calculation uses self-consistent shape mixing of the collective states, within the GCM framework, to obtain the NMEs (EDF$_{\text{full}}$).

Fig.~\ref{Fig3_EDF} shows that the $M^{0\nu}_{GT}$ pattern found with EDF spherical states disappears when PES minima are used.
Moreover, the NMEs are significantly reduced when the deformation effects are included.
Furthermore, the full EDF NMEs roughly follow the trends of the PES minima solution,
and configuration (shape) mixing only produces a shift to larger values,
which is larger in the Ti and Cr decays after the neutron $f_{7/2}$ orbital is filled.

On the other hand the SM calculations in Fig.~\ref{Fig3_SM} compare NMEs obtained using the KB3G interaction with seniority-zero initial and final states, the isospin projection of these states,
and the full pf calculation. Fig.~\ref{Fig3_SM} shows that isospin projection reduces significantly the Fermi component of the NME -panels (a), (c) and (e)- but is only a small correction to $M^{0\nu}_{GT}$ -panels (b), (d) and (f). For the GT component, the reduction is maximal at $N=Z$ nuclei, but very minor in the most neutron-rich systems.
The correlations associated to high-seniority components in the initial and final states are the responsible for the strong reduction of $M^{0\nu}_{GT}$, and these correlations also wash out the trend which appears with seniority-zero initial and final states.
In addition, it follows Figs.~\ref{Fig3_EDF}-~\ref{Fig3_SM} that these correlations reduce the NMEs more significantly than
the ones associated to collective deformation in the EDF approach.

\begin{figure}[h]
  \includegraphics[width=1.05\columnwidth]{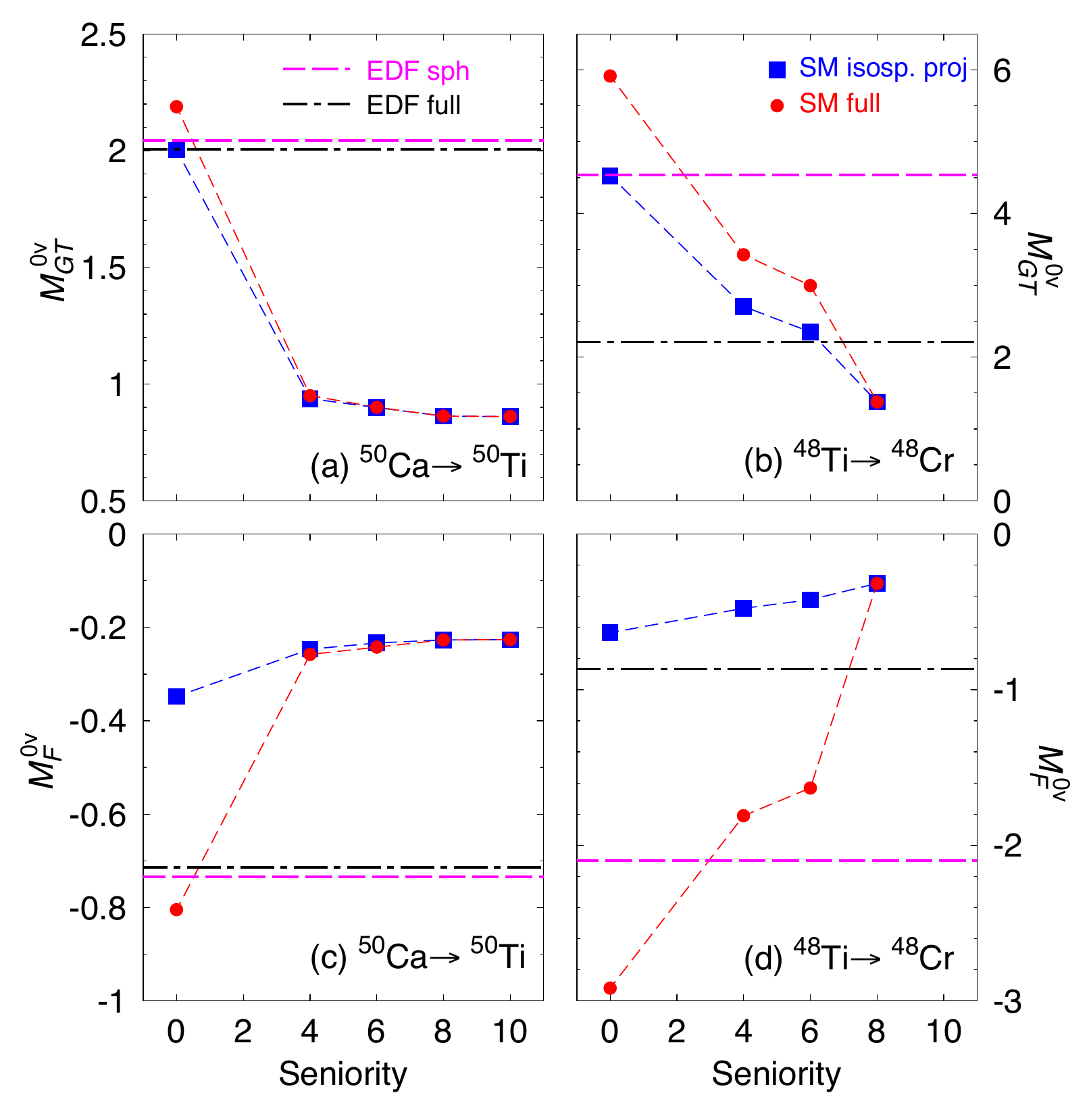}
\caption{(color online)
Gamow-Teller ($M^{0\nu}_{GT}$, panels a, b) and Fermi ($M^{0\nu}_{F}$, panels c, d) parts of the nuclear matrix element
of the non-physical $0\nu\beta\beta$ decays of $^{50}$Ca$\rightarrow^{50}$Ti (panels a, c) and $^{48}$Ti$\rightarrow^{48}$Cr (panels b, d).
Shell model (SM) results are shown as a function of the maximum seniority permitted in the initial and final states (squares),
and also after isospin projection(circles).
Energy density functional (EDF) results using spherical initial and final states (dashed lines) and the full EDF calculation (dashed-dotted lines) are also shown.
The EDF Gogny D1S and SM KB3G interactions are used.
}\label{isospin_nme} 
\end{figure}

Figure~\ref{isospin_nme} gives a detailed account of the evolution of the SM $M^{0\nu}_{GT}$ and $M^{0\nu}_{F}$ parts of the NMEs
as a function of the maximum seniority allowed in the initial and final nuclear states.
This figure shows that for the $^{50}$Ca$\rightarrow^{50}$Ti $0\nu\beta\beta$ decay, which relates two semi-magic nuclei,
seniority components up to $s=4$ are necessary for a reliable $M^{0\nu}_{GT}$ and $M^{0\nu}_{F}$ calculation.
The seniority decomposition of the full SM states is 97\%/3\%/0\% (77\%/21\%/2\%) for the $s=0$/$s=4$/$s>4$ components of $^{50}$Ca ($^{50}$Ti).
On the other hand, higher seniority components up to $s=8$ are needed in the $^{48}$Ti$\rightarrow^{48}$Cr decay. 
In this case the decomposition in seniority is 58\%/37\%/5\% (27\%/42\%/31\%) for the $s=0$/$s=4$/$s>4$ parts in $^{48}$Ti ($^{48}$Cr).
High-seniority components are therefore associated with the description of the deformed $^{48}$Cr.
Spherical and full EDF results are also shown in Fig.~\ref{isospin_nme}.
We have discussed above that spherical EDF results roughly correspond to seniority-zero SM calculations.
However, the full EDF NMEs behave quite differently in the two decays shown in in Fig.~\ref{isospin_nme}.
For $^{50}$Ca$\rightarrow^{50}$Ti decay, the final EDF number agree with the results of the spherical NME calculation.
This is due to the semi-magic character of the initial and final states,
which prevents any collective correlation in these nuclei (this also applies to the $^{42}$Ca$\rightarrow^{42}$Ti decay).
On the contrary, the full NMEs for the $^{48}$Ti$\rightarrow^{48}$Cr decay get contributions from collective deformation and shape mixing.
These final NMEs are roughly equivalent to the SM $s=6$ results.
This suggests that correlations associated to high-seniority components in the SM 
are not completely captured in EDF calculations.
These could be partially responsible for the differences between SM and EDF NMEs shown in Fig.~\ref{full_nme}.
Since the EDF states are built as linear combinations of projected Hartree-Fock-Bogoliubov-type states with different axial quadrupole deformations,
these intrinsic states are fully paired --in time-reversed single-particle orbits-- by definition.
Therefore, pair-breaking \textit{in the seniority scheme} is obtained by deforming the system,
but not by including explicitly quasiparticle excitations on top of each intrinsic state.
A step further, beyond the scope of this work, would include on equal footing both pair-breaking mechanisms into the GCM framework,
and study their influence in the NMEs.

Figure~\ref{isospin_nme} also shows that, when the seniority truncated $M^{0\nu}_{GT}$ results are projected to good isospin,
they are mildly reduced. On the contrary, isospin projection has an important effect for $M^{0\nu}_{F}$,
where states projected to good isospin are crucial.
Indeed, if the $r$ dependence of the neutrino potentials is removed,
the $M^{0\nu}_{F}$ connecting states with different $T$ values vanishes, as in two-neutrino double-beta decays.
In conclusion, only small changes in the GT part of the NME are expected from projecting EDF states to good isospin.
This also applies to other methods calculating NMEs which break isospin symmetry, such as the QRPA and IBM.


\section{Summary}~\label{summary}

We have studied the GT part of the NMEs of the Ca$\rightarrow$Ti, Ti$\rightarrow$Cr and Cr$\rightarrow$Fe $0\nu\beta\beta$ decays.
The systematic study of these non-physical decays allows us to compare shell-model and energy density functional calculations.
We observe that when full SM and EDF calculations are performed, SM results are about half the EDF values.
However, when we simplify the initial and final states of the decay
to spherical EDF and seniority-zero SM states, the NMEs obtained by both approaches are surprisingly similar,
suggesting that the nuclear structure description is equivalent for both methods at this level.
We have studied the role of nuclear structure correlations to the NMEs,
and we note that in general correlations associated to high-seniority SM components and EDF collective deformation
reduce the NMEs.
A comparison between these two suggests that the correlations associated to higher-seniority SM components
are not completely captured by the EDF approach, pointing to a possible reason for the difference between SM and EDF NMEs.
We have also explored projection to good isospin of the initial and final $0\nu\beta\beta$ decay states,
and conclude that, unlike for the Fermi part, it has only a moderate effect in the Gamow-Teller part of the NMEs.
This work opens up the door for benchmarks between NME calculated within different theoretical approaches,
and constitutes a step forward towards identifying the relevant ingredients that will lead to reliable NME calculations with reduced theoretical uncertainties. 

\section*{Acknowledgements}

This work was partly supported by the
Helmholtz Association through the Helmholtz Alliance Program, contract
HA216/EMMI ``Extremes of Density and Temperature: Cosmic Matter in the
Laboratory'', the Helmholtz International Center for
FAIR within the framework of the LOEWE program launched by the State of
Hesse, by the Deutsche Forschungsgemeinschaft through contract SFB~634 and by
the BMBF-Verbundforschungsprojekt number 06DA7047I.  
AP is partially supported by the MICINN (Spain) (FPA2011-29854); by the Comunidad de Madrid (Spain) (HEPHACOS S2009-ESP-1473) and by the European Union FP7 ITN INVISIBLES (Marie Curie Actions, PITN- GA-2011- 289442).

\end{document}